%% file: paper_fin.tex
\documentstyle[11pt,epsfig]{article}
\newcommand{\be}{\begin{equation}}
\newcommand{\ee}{\end{equation}}
\newcommand{\beqn}{\begin{eqnarray}}
\newcommand{\eeqn}{\end{eqnarray}}
\newcommand{\kf}{{\bf k}}
\newcommand{\lf}{{\bf l}}
\newcommand{\Pam}{I\!\!P}

\newcounter{savefig}
\newcommand{\alphfig}{\setcounter{savefig}{\value{figure}}%
\setcounter{figure}{0}%
\renewcommand{\thefigure}{\mbox{\arabic{savefig}\alph{figure}}}}

\begin{document}
DESY 96-026 

ANL-HEP-PR-96-11

hep-ph/9602363

\begin{center}
\begin{Large}
{\bf Quark-Antiquark Production in DIS Diffractive Dissociation}\\
\end{Large}
\vspace{0.5cm}
J. Bartels$^a$, H.Lotter$^a$,
                        M.W\"usthoff$^b$ \\
\vspace{0.5cm}
$^a${ II. Institut f\" ur Theoretische Physik, \\
Universit\" at Hamburg, Germany.}
\footnote
        {Supported by Bundesministerium f\"ur Forschung und
        Technologie, Bonn, Germany under Contract 05\,6HH93P(5) and
        EEC Program "Human Capital and Mobility" through Network
        "Physics at High Energy Colliders" under Contract
        CHRX-CT93-0357 (DG12 COMA).}
\\
$^b${High Energy Physics Division, \\
Argonne National Laboratory, USA.}
\footnote{Supported by the U.S. Department of Energy, 
Division of High Energy Physics,
Contract W-31-109-ENG-38.}

\end{center}
\vspace{2.0cm}
\begin{abstract} 
\noindent
We calculate the cross section for the production of two jets with large 
transverse momenta $\kf^2$ in DIS 
diffractive dissociation for both transverse and longitudinally polarized
photons. The scale which defines the hardness of the Pomeron is found to be
$\kf^2 \frac{Q^2+M^2}{M^2}$. We present analytic expressions and discuss 
numerical results relevant for the diffractive dissociation at HERA.
\end{abstract}
\vspace{2cm}
{\bf 1.)}
The diffractive dissociation of the photon in deep inelastic ep-scattering
at HERA has recently attracted much interest. As for the inclusive diffractive
cross section attention has been given to the question whether the Pomeron
which describes the energy dependence of this process is closer to the soft
Pomeron which describes the high energy behavior of hadron-hadron scattering
or whether 
it resembles the hard Pomeron observed in the proton structure function at 
small $x$. Whereas earlier analysis \cite{h11,zeus1} of HERA data seem to 
favor the first alternative, a recent analysis \cite{zeus2} presents 
evidence for a strong admixture of the hard Pomeron. A natural way to gain
further clarification is a more detailed analysis of the diffractive
final state. Electroproduction of vector mesons and photoproduction of 
$J/\Psi$
have been analysed \cite{jpsi1} 
and compared \cite{jpsi2} 
with both the soft and the hard Pomeron, and
they seem to favor a harder Pomeron. A new class of final states which allow
to enhance the hard Pomeron component are jets with large transverse momenta
\cite{r,d,nnn}. The simplest configuration are two-jet events with the jets
originating from $q\bar{q}$ pairs; for larger invariant masses $M$ one expects
events with additional production of gluon jets to become dominant.

In this paper we consider the simpler case $q\bar{q}$ jet production, 
restricting ourselves to the kinematic region where the hard Pomeron can be
expected to dominate (fig.\ 1). Starting from the DIS diffractive cross section
formula
\beqn
\frac{d\sigma^{ep}}{dydQ^2dtdM^2d\kf^2}\;=\;\frac{\alpha_{em}}{yQ^2\pi}\left[
\frac{1+(1-y)^2}{2}\frac{d\sigma_T^{\gamma^* p}}{dtdM^2d\kf^2}\;+\;
(1-y)\frac{d\sigma_L^{\gamma^* p}}{dtdM^2d\kf^2}\right]
\eeqn
we calculate the differential cross section for the subprocess $\gamma^*+p
\to q\bar{q} +p$ for very small $t$ and consider both the transverse 
and longitudinal cross section.  
\\ \\
{\bf 2.)} 
We assume the total energy $s=(p+q)^2$ 
to be much larger than the photon virtuality $Q^2=-q^2$ 
as well as the invariant mass of the quark anti-quark pair 
$M^2=(q+x_{\Pam}p)^2$, i.e. the Pomeron momentum is much smaller than
the proton momentum ($x_{\Pam} \ll  1$). 
Instead of $M^2$ one can also introduce 
the variable $\beta$ which is defined as $\beta=Q^2/(M^2+Q^2)$.
For our analytic calculation we set the momentum transfer $t$ to zero,
since the cross section strongly peaks at $t=0$. In order to compare
with data we will add the $t$-dependence later on by hand. 
 
The new element in our approach is the fixed transverse 
momentum $\kf$ of the (anti-)quark which enters as a second hard scale 
in our calculation and allows to treat the Pomeron perturbatively 
(hard Pomeron). 
For large $s$ (small $x_{\Pam}$) the
amplitude of the process is dominated by two-gluon
exchange, and following the $\kf$-factorization theorem
\cite{cath} we express the amplitude with leading-log($1/x_{\Pam}$) 
accuracy through the unintegrated gluon distribution
of the proton. The two gluons have to couple to the 
outgoing quarks in all possible combinations in order to preserve
gauge invariance. \\ 
The quark phase space is usually parameterized in terms of 
the transverse momentum $\kf$ and the 
longitudinal momentum fraction $\alpha$ according to the Sudakov
decomposition: $k+x_{\Pam}p=\alpha q'+ \kf^2/(\alpha s) p +\kf$ 
(the anti-quark momentum consequently has the parametrization:
$q-k=(1-\alpha) q'+ \kf^2/[(1-\alpha)s] p -\kf$) with
$q'=q+xp$. The invariant mass can now easily be expressed in terms of
$\alpha$ and $\kf^2$: $M^2=2(q-k)\cdot(k+x_{\Pam}p)=\kf^2/[\alpha(1-\alpha)]$.
In the quark anti-quark CMS one finds for $\alpha=(1-\cos(\theta))/2$ where
$\theta$ denotes the angle between the quark and the proton. From the 
expression $\kf^2=\alpha (1-\alpha)M^2$  it becomes clear that keeping 
the mass $M$ fixed (of the order
of $Q$) and taking $\kf^2$ to be small also means that $\alpha$
and $\theta$ become small. As will be seen below 
the cross section is dominated by small $\kf^2$, i.e.\ the jets appear most
frequently at small angles or phrased in different words the jets are 
'aligned' (see Aligned Jet Model \cite{bj}). 

The cross section for the $q\bar{q}$ production has the following form
 (for details see refs. \cite{d,m,nz,bw,wue,lw}):
\beqn
\frac{d \sigma_T}{d M^2 dt d\kf^2}_{|t=0}&=&
\sum_f e_f^2 \frac{\alpha_{em}\pi^2\alpha_s^2}{12} \frac{1}{M^4}
\frac{(1-\frac{2\kf^2}{M^2})}{\sqrt{1-\frac{4\kf^2}{M^2}}}
\left[I_T(Q^2,M^2,\kf^2)\right]^2 
\label{sig1t}
\\
\frac{d \sigma_L}{d M^2 dt d\kf^2}_{|t=0}&=&
\sum_f e_f^2 \frac{\alpha_{em}\pi^2\alpha_s^2}{3} \frac{4}{Q^2M^2}
\frac{\kf^2}{M^2} \frac{1}{\sqrt{1-\frac{4\kf^2}{M^2}}}
\left[I_L(Q^2,M^2,\kf^2)\right]^2 
\label{sig1l}
\eeqn
with
\beqn
I_T(Q^2,M^2,\kf^2)&=&
-\int  \frac{d\lf^2}{\lf^2}{\cal F}_G(x_{\Pam},\lf^2)\left[
\frac{M^2-Q^2}{M^2+Q^2}+
\frac{\lf^2+\frac{\kf^2}{M^2}(Q^2-M^2)}{
\sqrt{(\lf^2+\frac{\kf^2}{M^2}(Q^2-M^2))^2
+4\kf^4\frac{Q^2}{M^2}}}\right]
\label{ktt}
\\
I_L(Q^2,M^2,\kf^2)&=&
-\int  \frac{d\lf^2}{\lf^2}{\cal F}_G(x_{\Pam},\lf^2)\left[
\frac{Q^2}{(M^2+Q^2)}-\frac{\kf^2\; Q^2}{M^2
\sqrt{(\lf^2+\frac{\kf^2}{M^2}(Q^2-M^2))^2
+4\kf^4\frac{Q^2}{M^2}}}\right] 
\label{ktl}
\eeqn
where ${\cal F}_G$ represents the unintegrated gluon distribution
of the proton
\beqn
\int^{Q^2} d \lf^2  {\cal F}_G(x_{\Pam},\lf^2) 
=x_{\Pam}G(x_{\Pam},Q^2)\;\;. 
\label{glue}
\eeqn
This identification of the Pomeron with the 
gluon structure function, however, has to be taken with care. 
Strictly speaking the function which appears in eq.\ 
(\ref{glue}) is not exactly 
the same gluon structure function as in DIS. Namely the longitudinal 
component of the two gluon lines in fig.\ 1, even at $t=0$, 
have a typical $\beta$-value of the order 
of $x_{\Pam}$ and their difference is exactly $x_{\Pam}$. In the 
leading-log($1/x_{\Pam}$) approximation where we cannot distinguish 
log($1/x_{\Pam})$ and log($2/x_{\Pam}$) 
the function $G$ from (\ref{glue}) has the same 
$x_{\Pam}$-dependence as the gluon structure 
function. But in the HERA-regime we cannot exclude a difference
in the absolute normalization. In the following we shall identify our function 
$G$ with the gluon structure function, but in a more ambitious analysis
one could think of calculating the nonforward gluon structure function using 
the generalized gluon splitting function of \cite{bfkl2}. 

Note the extra $Q^2$ in the denominator of the longitudinal 
cross section in (3) which reflects the fact that 
this is a higher twist contribution. The term 
$[...]/\lf^2$ in the integrand of eq.\ (\ref{ktt}) 
is roughly constant for $\lf^2$ smaller than 
$\kf^2(M^2+Q^2)/M^2$ and falls off like $1/\lf^2$ for larger $\lf^2$.
The unintegrated gluon structure function ${\cal F}_G$ also
falls like $1/\lf^2$ (modified by logarithms) so that the dominant 
contribution in the integration 
comes from the region $\lf^2<\kf^2(M^2+Q^2)/M^2$. We approximate 
the integrand here by taking the limit $\lf^2 = 0$ for $[...]/\lf^2$ 
and integrate over ${\cal F}_G$ with $\kf^2(M^2+Q^2)/M^2$ 
as the upper limit. The latter results in 
$x_{\Pam}G(x_{\Pam},\kf^2(M^2+Q^2)/M^2)$,
i.e.\ we have extracted the leading log($\kf^2(M^2+Q^2)/M^2$) contribution.
The caveat, however, is that this approximation underestimates the region
of large $M^2$. This becomes clear by first taking the limit $M^2 \rightarrow
\infty$ and then performing the integration over $\lf^2$.
In this limit $[...]$ equals $2\theta(\lf^2-\kf^2)$, i.e.\ 
there is no contribution at low $\lf^2$. The integrand at large $\lf^2$ is 
roughly falling like $1/\lf^4$ and 
the integral is determined by its lower limit
$\kf^2$ which approximately leads to ${\cal F}_G(x_{\Pam},
\kf^2)/\kf^2$, i.e.\ the final result also contains
the unintegrated structure function. In the next section we present the 
correction for any $M$-value which unfortunately cannot be derived from simple
arguments as given here.
Physically the scale $\kf^2(M^2+Q^2)/M^2$ corresponds to the virtuality of the 
softer of the two quarks, into which the photon dissociates.
With the approximation described above one obtains:
\beqn
I_T&=&
\left[ \frac{4Q^2M^4}{\kf^2(M^2+Q^2)^3}\;+\;
b_t\frac{\partial}{\partial \kf^2}
\right] \; x_{\Pam}G(x_{\Pam},\kf^2 \frac{Q^2+M^2}{M^2})
\label{dlat}
\\
I_L&=&\left[
\frac{Q^2M^2(Q^2-M^2)}{\kf^2(M^2+Q^2)^3}\;+\;b_l
\frac{\partial}{\partial \kf^2} 
\right]x_{\Pam}G(x_{\Pam},\kf^2 \frac{Q^2+M^2}{M^2}).
\label{dlal}
\eeqn
where $b_t$ and $b_l$ are functions of $M^2$ and $Q^2$ and will be given 
in (11) and (12). Let us discuss the main properties of this 
result. It is seen that the transverse
momentum of the outgoing quark pair sets the scale of
the gluon structure function. This means that the effective slope
of the hard Pomeron, which determines the rise of the 
cross section at small $x_{\Pam}$ should increase with
increasing transverse momentum. The effect is even strengthened
by the additional factor $(Q^2+M^2)/M^2$ which enters the scale
and is of the order of two for the $Q^2,M^2$-range considered here. From
the known behavior of the gluon structure function
we can furthermore conclude that the power rise of the 
cross section as a function of $\kf^2$ is damped for 
$\kf^2$ small, but still in a range where the perturbative 
approach is legitimate. Note that (7) and (8) cannot be extrapolated
down to $\kf^2=0$. 
The fall-off with increasing $M^2$ is identical for 
both polarizations.
A remarkable property of the first term of the longitudinal cross section 
in (8) is its zero for $Q^2=M^2$.
A numerical evaluation of the above formulas will be performed 
in part {\bf 4.)}.
\\ \\
{\bf 3.)}
Thus far we have discussed the cross section in the
double-leading-log approximation. One can in principle do
better because the $\kf$-factorization formulas eq.\ (\ref{ktt}),
(\ref{ktl}) are more general and allow for an evaluation
without the leading-log($Q^2$)-approximation.
This is equivalent to using the BFKL Pomeron \cite{bfkl} for the 
unintegrated gluon structure function.
We will show that the 
characteristic momentum scale which was identified
in the previous section emerges in this model
without any approximation and that we essentially 
reproduce the formulas given there, provided this 
scale is large enough.  

The use of the BFKL Pomeron requires a 
nonperturbative input distribution which cannot be determined
theoretically. Using only the most general
parameterization we find for the 
BFKL Pomeron the following expression
\beqn
{\cal F_G}(x_{\Pam},\lf^2)\; =\; \frac{1}{\Lambda_0^2}\int_{-\infty}^{+\infty}
\frac{d \nu}{2 \pi}\left(\frac{\lf^2}{\Lambda_0^2}
\right)^{-\frac{1}{2}-i\nu} 
 \phi(\nu)\;\exp\left[ \chi(\nu) \log\frac{1}{x_{\Pam}}
\right]
\eeqn
with $\Lambda_0^2$ being a nonperturbative scale ($\Lambda_0^2 
\simeq 1 \mbox{GeV}^2$), $\phi(\nu)$ a slowly varying function of 
$\nu$ and $\chi(\nu)=\frac{N_c\alpha_s}{\pi}
[2\psi(1)-\psi(1/2+i\nu)-\psi(1/2-i\nu)]$ 
the characteristic eigenfunction of the BFKL pomeron. 
Now we insert the above expression in our basic equations 
(\ref{ktt}), (\ref{ktl}) and perform the $\lf^2$-integration
exactly. This leads to
\beqn
I_T&=&\frac{2\kf^2}{\Lambda_0^4}\int_{-\infty}^{+\infty}\frac{d \nu}{2 \pi}  
\left[\frac{\Lambda_0^2M^2}{\kf^2(Q^2+M^2)}\right]^{\frac{3}{2}+i\nu}
\phi(\nu)\;\exp\left[ \chi(\nu) \log\frac{1}{x_{\Pam}}\right]
\;\;\cdot \nonumber\\  &&\hspace{2cm}\cdot\;\;
(\frac{3}{2}+i\nu)\Gamma(\frac{1}{2}+i\nu)\Gamma(\frac{1}{2}-i\nu)
F(\frac{3}{2}+i\nu,-\frac{1}{2}-i\nu,2;\frac{M^2}{M^2+Q^2})
\\
I_L&=&\frac{\kf^2 Q^2}{M^2\Lambda_0^4}
\int_{-\infty}^{+\infty}\frac{d \nu}{2 \pi}  
\left[\frac{\Lambda_0^2M^2}{\kf^2(Q^2+M^2)}\right]^{\frac{3}{2}+i\nu}
\phi(\nu)\;\exp\left[ \chi(\nu) \log\frac{1}{x_{\Pam}}\right]
\;\;\cdot \nonumber \\  &&\hspace{2cm}\cdot\;\;
\Gamma(\frac{1}{2}+i\nu)\Gamma(\frac{1}{2}-i\nu)
F(\frac{3}{2}+i\nu,-\frac{1}{2}-i\nu,1;\frac{M^2}{M^2+Q^2}),
\eeqn
where $F$ is the hypergeometric function.
The remaining $\nu$-integral can be safely evaluated numerically
but here we restrict ourselves to the discussion of some limiting cases.
One observes immediately that the scale $\kf^2(Q^2+M^2)/M^2$ is the 
relevant parameter, which determines the location of the saddle point
of the integral.  As long as $N_c \alpha_s / \pi \cdot
\log 1/x_{\Pam}$ is much larger
than $\Delta=\log [\kf^2(Q^2+M^2)/(M^2\Lambda_0^2)]$ we are in the
BFKL-limit and the saddle point lies at $\nu \simeq 0$.
In this case one finds the usual BFKL results with a steep rise 
of the cross section for small $x_{\Pam}$. 

If, on the other hand, $\Delta$ is much larger than 
$\ln 1/x_{\Pam}$, we are in the double logarithmic
limit and the $\nu$-contour has to be shifted close to $\nu=-i/2$.
The saddle point is located at 
$i\nu_s=1/2-\sqrt{N_c\alpha_s/(\pi \Delta) \ln (1/x_{\Pam})}$.
Evaluating the $\nu$-integral we obtain
\beqn
\frac{d \sigma_T}{d M^2 dt d\kf^2}_{|t=0}&=&
\sum_f e_f^2 \frac{\alpha_{em}\pi^2\alpha_s^2}{3} 
\frac{(1-\frac{2\kf^2}{M^2})}{\sqrt{1-\frac{4\kf^2}{M^2}}}
\frac{M^4}{(M^2+Q^2)^4}\;\left[
\left\{\frac{2Q^2}{\kf^2(M^2+Q^2)}\right.\right.
\nonumber\\ &+&\;\left.\left.
\left[\frac{M^2-Q^2}{M^2+Q^2}-\frac{2Q^2}{M^2+Q^2}\ln\left(\frac{Q^2}{M^2+Q^2}
\right)\right]\frac{\partial}{\partial \kf^2}
\right\}\; x_{\Pam}G(x_{\Pam},\kf^2 \frac{Q^2+M^2}{M^2})\right]^2
\label{fint}
\\
\frac{d \sigma_L}{d M^2 dt d\kf^2}_{|t=0}&=&
\sum_f e_f^2 \frac{\alpha_{em}\pi^2\alpha_s^2}{3}
\frac{4}{\sqrt{1-\frac{4\kf^2}{M^2}}}
\frac{\kf^2Q^2}{(M^2+Q^2)^4}\;\left[
\left\{\frac{Q^2-M^2}{\kf^2(M^2+Q^2)}\right.\right.
\nonumber\\ &+&\;\left.\left.
\left[\frac{2M^2}{M^2+Q^2} +\frac{M^2-Q^2}{M^2+Q^2}
\ln\left(\frac{Q^2}{M^2+Q^2}\right)\right]\frac{\partial}{\partial \kf^2}
\right\}\;x_{\Pam}G(x_{\Pam}, \kf^2 \frac{Q^2+M^2}{M^2})\right]^2
\label{finl}
\eeqn
with the double leading log approximation of eq.\ (\ref{glue})
\beqn
x_{\Pam}G(x_{\Pam},\kf^2 \frac{Q^2+M^2}{M^2})\;=\;
\frac{\left[N_c\alpha_s/(\pi \Delta) \ln (1/x_{\Pam})\right]^{-1/4}}
{\sqrt{8N_c\alpha_s}}\;
\exp\left(\sqrt{4 \frac{N_c\alpha_s}{\pi} 
\ln\left(\frac{1}{x_{\Pam}}\right)\Delta}\right) 
\phi(\nu_s).
\label{dlaglue}
\eeqn
If we compare these results with those obtained in the
previous section, we see that an improvement has been achieved
in the prefactor of the unintegrated structure function.
It is more accurate in the transverse case and we have found a correction 
for the longitudinal part, too, where the zero for $Q^2=M^2$ disappears. 
The leading term, however, which contains the structure function itself
is the same as before. 
We should stress here again that the scale of the gluon 
structure function $\kf^2 (Q^2+M^2)/M^2$ already 
appeared in the single 
logarithmic approximation and turned out to be the 
most important parameter in this process. 

Finally, let us comment on the region of small $\kf^2$. The expressions given
in (10) and (11) can be continued down to $\kf^2=0$ and the integral
gives a finite answer. For the transverse cross section a detailed saddle 
point analysis shows that the main contribution comes from the small 
$\kf^2$-region and thus agrees with the Aligned Jet Model. For the 
longitudinal cross section on the other hand, the additional factor $\kf^2$
makes the small $\kf^2$-region less dominant. Hence, for the 
$\kf^2$-integrated transverse cross section the BFKL model gets a too big 
nonperturbative contribution and thus is not reliable. The longitudinal
cross section has less nonperturbative contributions.

The situation becomes better, if $\kf^2$ is constrained to be large.
This was shown for 
diffractive jet production in a numerical study 
contained in \cite{blv}. From
this we draw the 
conclusion that our above analysis, for both cases, is on a 
solid theoretical basis if only $\Delta$, which is
the logarithm of the relevant scale here, is large.
Hence we regard the  eqs.\ (\ref{fint}), (\ref{finl})
as a firm prediction,
up to a normalization uncertainty,
in a range, where $\kf^2$ is at least 
of the order of $1 \mbox{GeV}^2$ and where the rise of the gluon density
as a function of $x$ is visible, the latter condition 
ensuring that we are in the small-$x$ limit.
\\ \\
{\bf 4.)} Based upon the formulae 
(\ref{sig1t}), (\ref{sig1l}), (\ref{fint}) and 
(\ref{finl}) we
have performed a few numerical calculations which may apply to HERA data.
Throughout our calculations we have integrated over $t$ by taking 
our cross section expression evaluated at $t=0$ and multiplying
with the formfactor which was given by
Donnachie and Landshoff \cite{dl}.   
Furthermore we used the running strong coupling constant with 
the scale being given by the scale of the gluon structure function.
First we have looked into the most striking feature of the cross section
formula, the rise of the gluon structure function at small $x_{\Pam}$.
Fig.\ 2a and 2b 
show the $x_{\Pam}$-dependence of the transverse and the longitudinal cross
sections, keeping $\beta$ and $Q^2$ fixed at $2/3$ and $50\mbox{GeV}^2$,
respectively, and 
integrating over $\kf^2 > 2\mbox{GeV}^2$.
We have considered different 
parameterizations for the gluon structure function: the 
GRV leading order and
next-to-leading-order parameterizations
\cite{grv}, and the hybrid model of  
one of us (M.W.) \cite{wue}. For comparison we also show a 
prediction of the soft Pomeron \cite{d}. As expected, the hard 
Pomeron rises much stronger at small $x_{\Pam}$ than the soft Pomeron.
The upper three curves show the overall uncertainty in both
$x_{\Pam}$-shape and overall normalization 
of the perturbative approach. In order to find the ``optimal``
version we have calculated 
the inclusive structure function
$F_2$, by coupling the gluon to the quark loop
and doing the same type of approximations as for the $q\bar{q}$ cross section
described in the previous 
section (here we have assumed that in the kinematic region
which we are considering the gluon structure function dominates). 
Figs.\ 3a-d show the results of this computation in comparison
with data \cite{data}, and
we feel that the next-to-leading order GRV parameterization gives the best 
description. After having in this way ``selected`` which parameterization 
for
the gluon structure function is the best one for our purposes, we return 
to eqs. (\ref{fint}) and (\ref{finl}) 
and calculate the $x_{\Pam}$-dependence, varying now the 
momentum scale ${\bf k^2} (Q^2+M^2)/M^2$ which determines the
``hardness` of the Pomeron structure function. As it can be seen from
fig.\ 2c,
there is a clear variation in the slope (numerical values for the slopes
are given in the figure), but in the HERA range this variation is not
very strong. So the main experimental signal, for the beginning, may be the
observation of a rise in eqs.\ 
(\ref{fint}) and (\ref{finl}) with a power of $1/x_{\Pam}$ 
around 0.75.

Next we have used our formulae for estimating the ep-integrated cross section
for the production of two jets with large transverse momenta. We define two
cuts; (a) $ \kf^2 > 2 \mbox{GeV}^2$
 and (b) $\kf^2 > 5 \mbox{GeV}^2$, and we then
integrate over all other variables in the region
 $Q^2 > 10 \mbox{GeV}^2$, $x_{\Pam} < 10^{-2}$ and 
$50 \mbox{GeV} < W < 220\mbox{GeV}$ ($W^2=(p+q)^2$). The results for
the transverse and the longitudinal cross sections are given in Table 1.
Typically the longitudinal cross section is of the order $10\%$ of the
transverse cross section. The other conclusion that may be drawn from this
table concerns the dependence upon the $\kf^2$ cut: moving from the cut 
(a) to cut (b), one loses about $75 \%$ of the cross section.
%%%%%
\begin{table}[h]
\begin{center}
\begin{tabular}{|c|c|c|c|c|}   \hline
 & $\kf^2_0 = 2 \mbox{GeV}^2$& & $\kf^2_0 = 5 \mbox{GeV}^2$ & \\
 \hline \hline
               &  GRV(LO)  &  GRV(NLO) & GRV(LO)   & GRV(NLO) \\
 \hline \hline
 $\sigma^{ep}_T$    & 193     & 108     & 29    & 19  \\
  \hline
 $\sigma^{ep}_L$    & 15     &  9    &  2   & 1  \\
  \hline
 $\sum_{i=T,L}\sigma^{ep}_i$
                    &  208    &  117    & 31    & 20 \\
  \hline
\end{tabular}
\end{center}
\caption{Results for total ep-cross sections (in pbarn) of 
diffractive dijet production for two 
different parameterizations of the gluon density and 
two different cuts on the transverse momentum of the jets.}
\end{table}
%%%%%%
In fig.\ 4 we show the dependence on $\kf^2$, keeping $\beta$ and $Q^2$
fixed. On the whole one notices that the longitudinal cross
section (figs.\ 4 b,d)
decreases less rapidly than the transverse one (figs.\ 4 a,c).
According to our formulae, the transverse cross section falls approximately 
as $1/\kf^4$, whereas the longitudinal one decreases more like 
$1/\kf^2$. This simple power behavior is, however, slightly
modified by the $\kf^2$-dependence of the gluon structure function 
(numerical values for the slopes are given in the figures).
In the region of
smaller $\kf^2$ the cross sections start to become somewhat flatter.
This is in agreement with what we have already discussed  
after eq.\ (\ref{dlal}). The most
striking variation is seen in fig.\ 4b: in the
double leading log approximation
the longitudinal cross section has a zero at $Q^2=M^2$: due to the correction
term in (\ref{finl}), 
this zero is somewhat shifted towards smaller $\beta$. 
What we see in fig.\ 4b is the remainder of this zero.

Another quantity of interest is the $\beta$-spectrum for jets with $\kf^2$
larger than 
$2 \mbox{GeV}^2$ (fig.\ 5). As to the transverse cross section, 
in comparison with
the inclusive diffractive cross for $q\bar{q}$ production which shows the
$(1-\beta)$-dependence we now see a maximum which at not too large $Q^2$ lies
below $\beta=0.5$. In the longitudinal cross section which 
again is smaller than
the transverse cross section by a factor of 10 we see the dip which
has moved from the value $\beta=1/2$ (uncorrected cross section)
to $\beta=0.4$ or less.
\\ \\
{\bf 5.)} 
In this paper we have collected evidence that the observation of two jets with
high transverse momentum  may provide a valuable way of analysing, in the DIS
Diffractive Dissociation, the hard Pomeron. After having established that
the scale for the hardness of the Pomeron is not simply $\kf^2$ but
the combination $\kf^2(Q^2+M^2)/M^2$ we have illustrated several
signals of this hard Pomeron. In our numerical estimates we have found that
it is not enough to consider that part of the phase space where the
transverse momenta are strongly ordered.
The main signal of the hard Pomeron is, 
most likely, the stronger rise in $1/x_{\Pam}$.
But there are also other, more
subtle details which are typical for the hard Pomeron. A more careful
comparison with the predictions of the soft Pomeron seems very important.

Production of two jets clearly represents the simplest case of jet production 
in DIS Diffractive Scattering. In the next step one has to compute
higher order corrections to two jet production and to generalize to the 
production of of extra gluon jets which are expected to become important 
especially in the large $M$-region. Another point of interest is the 
$t$-dependence. Whereas the soft Pomeron predicts the typical 
shrinkage for the $t$-slope, the hard Pomeron is expected, at least for $t$ 
of the order of 
$1 \mbox{GeV}^2$, to have a much weaker energy dependence of the $t$-slope.
Both these issues have to be analysed in more detail in order to 
reach a more complete understanding of the hard Pomeron in Diffractive
DIS.
\\ \\
{\bf Acknowledgements:} We thank M.Diehl for useful discussions and for 
his help
in preparing figs.\ 2 a,b. We gratefully acknowledge the help of C.Ewerz in
comparing our $F_2$ results with the experimental data.\\ \\
%
%%%%%%%%%%%%%%%%%%%%%%%%%%%%%%%%
%%%%%%%%%%%%%%%%%%%%%%%%%%%%%%%% 

%%%%%%%%%%%%%%%%%%%%%%%
\newpage
%%%%%%%%%%%%%%%%%%%%%%%
\section*{Figure Captions}
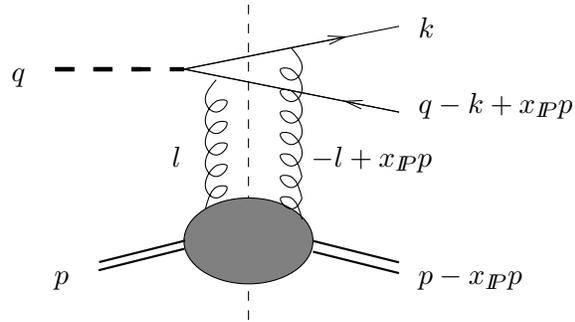
\begin{figure}[htbp]
\begin{center}
\input fig1.pstex_t
\end{center}
\caption{
\label{f1}}
\end{figure}

\begin{description}
\item
Fig.\ \ref{f1} : 
A contribution to the amplitude for the process 
$\gamma^*+p \to  q\bar{q}+p$. The shaded blob represents the unintegrated
gluon structure function. For the full amplitude the gluons have to 
be coupled to the quarks in all possible ways.
\\
\item
Fig.\ \ref{f2a} : 
The $x_{\Pam}$-dependence of the cross section $d \sigma ^{\gamma^* p}/
d \beta$ for transverse photons. The curves are for $Q^2=50 \mbox{GeV}^2$,
$\beta=2/3$ and $\kf^2$ integrated from $2 \mbox{GeV}^2$ up to the kinematical 
limit. Results are shown for the soft Pomeron exchange model of 
\cite{d}, the perturbative hybrid model of \cite{wue} 
and the model presented in this paper with the GRV leading order
and next-to-leading order gluon distribution \cite{grv}.
\\
\item
Fig.\ \ref{f2b} : 
The same as fig.\ \ref{f2a} but for longitudinal photons.
\\
\item
Fig.\ \ref{f2c} : 
The $x_{\Pam}$-dependence of the cross section 
$1/N \cdot d \sigma ^{\gamma^* p}/
d \beta$ for transverse photons using eq.\ (\ref{fint}) and the 
GRV(NLO)-parameterization for three different kinematical
situations: $Q^2=80 \mbox{GeV}^2$, $\beta=2/3$, 
$\kf^2$ integrated from $2\mbox{GeV}^2$ to $4\mbox{GeV}^2$ (solid line),
$Q^2=80 \mbox{GeV}^2$, $\beta=2/3$,   
$\kf^2$ integrated from $4\mbox{GeV}^2$ to $8\mbox{GeV}^2$ (dotted line)  
and
$Q^2=80 \mbox{GeV}^2$, $\beta=5/6$,   
$\kf^2$ integrated from $2\mbox{GeV}^2$ to $4\mbox{GeV}^2$ (dashed line).
The numbers in the figure give the slope of each curve.  
The normalization $N$ is the integral of the cross section over the
$x_{\Pam}$-range displayed.
\\
\item
Figs.\ \ref{f3a}-\ref{f3d} : 
Comparison of our results for the inclusive structure function $F_2$,
calculated with the same type of approximation as described in section
{\bf 3.)} and two different parameterizations of the gluon density
\cite{grv}
with data from 
H1 , ZEUS and E665 \cite{data} for different values of $Q^2$.
\\
%%\item
%%Fig.\ \ref{f3b} : 
%%\\
%%\item
%%Fig.\ \ref{f3c} : 
%%\\
%%\item
%%Fig.\ \ref{f3d} : 
%%\\
\item
Fig.\ \ref{f4a} : 
The $\kf^2$-dependence of the differential $\gamma^* \,p$-cross section
for transverse photons according to eq.\ (\ref{fint}). 
Values for the kinematical parameters are $x_{\Pam}=5 \cdot 10^{-3}$,
$\beta=2/3$, and $Q^2$ is varied between $15 \mbox{GeV}^2$ and 
$45 \mbox{GeV}^2$. 
The quantity $\delta$ gives the effective slope of the curves,
obtained from a numerical fit to a power 
behaviour $\sim (\kf^2)^{-\delta}$.
\\
\item
Fig.\ \ref{f4b} : 
The same as in fig.\ \ref{f4a} but for longitudinal 
photons (eq.\ (\ref{finl})).
\\
\item
Fig.\ \ref{f4c} :
The same as in fig.\ \ref{f4a} but for fixed $Q^2=10\mbox{GeV}^2$
and different values of $\beta$. 
\\
\item
Fig.\ \ref{f4d} : 
The same as in fig.\ \ref{f4c} but for longitudinal photons.
\\
\item
Fig.\ \ref{f5a} : 
The $\beta$-dependence of the cross section 
$d \sigma ^{\gamma^* p}/
d \beta$ for transverse photons. The curves are for       
$x_{\Pam}=5 \cdot 10^{-3}$ and three different values of $Q^2$.
\\
\item
Fig.\ \ref{f5b} : 
The same as in fig.\ \ref{f5a} but for longitudinal photons.
\end{description}
%%%%%%%%%%%%%%%%%%%%%%%
%%%%%%%%%%%%%%%%%%%%%%%
\section*{Figures}
%%%
\setcounter{figure}{2}
\alphfig
\begin{figure}[htbp]
\begin{center}
\input fig2a.pstex_t
\end{center}
\caption{
\label{f2a}}
\end{figure}
%%%
\begin{figure}[htbp]
\begin{center}
\input fig2b.pstex_t
\end{center}
\caption{
\label{f2b}}
\end{figure}
%%%
\begin{figure}[htbp]
\begin{center}
\input fig2c.pstex_t
\end{center}
\caption{
\label{f2c}}
\end{figure}
\setcounter{figure}{3}
\alphfig
\begin{figure}[htbp]
\begin{center}
\input at2.pstex_t
\end{center}
\caption{
\label{f3a}}
\end{figure}
%%%%
\begin{figure}[htbp]
\begin{center}
\input at4.5.pstex_t
\end{center}
\caption{
\label{f3b}}
\end{figure}
%%%
\begin{figure}[htbp]
\begin{center}
\input at8.5.pstex_t
\end{center}
\caption{
\label{f3c}}
\end{figure}
\begin{figure}[htbp]
\begin{center}
\input at15.pstex_t
\end{center}
\caption{
\label{f3d}}
\end{figure}
%%%
\setcounter{figure}{4}
\alphfig
\begin{figure}[htbp]
\begin{center}
\input fig4a.pstex_t
\end{center}
\caption{
\label{f4a}}
\end{figure}
\begin{figure}[htbp]
\begin{center}
\input fig4b.pstex_t
\end{center}
\caption{
\label{f4b}}
\end{figure}
%%%
\begin{figure}[htbp]
\begin{center}
\input fig4c.pstex_t
\end{center}
\caption{
\label{f4c}}
\end{figure}
\begin{figure}[htbp]
\begin{center}
\input fig4d.pstex_t
\end{center}
\caption{
\label{f4d}}
\end{figure}
%%%
\setcounter{figure}{5}
\alphfig
\begin{figure}[htbp]
\begin{center}
\input fig5a.pstex_t
\end{center}
\caption{
\label{f5a}}
\end{figure}
\begin{figure}[htbp]
\begin{center}
\input fig5b.pstex_t
\end{center}
\caption{
\label{f5b}}
\end{figure}
%%%
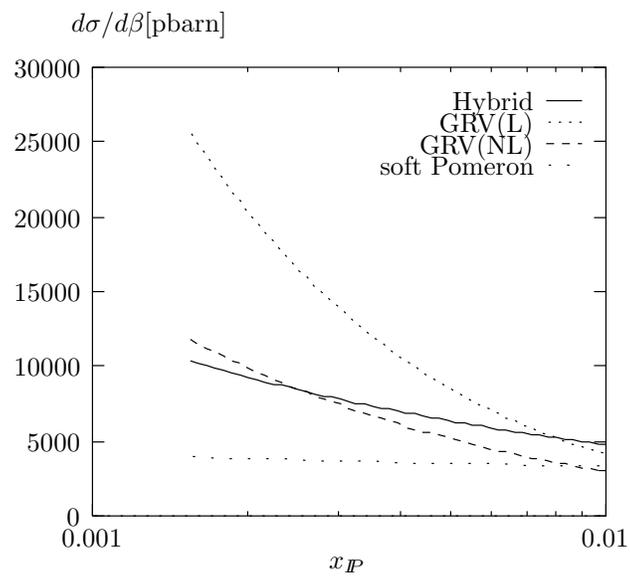
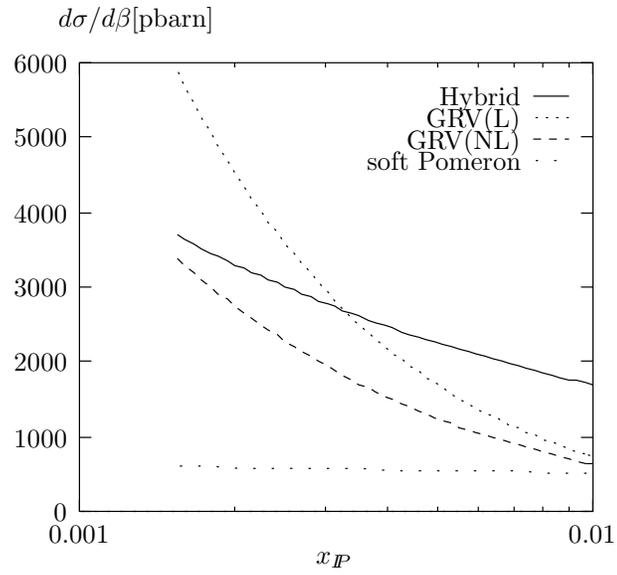
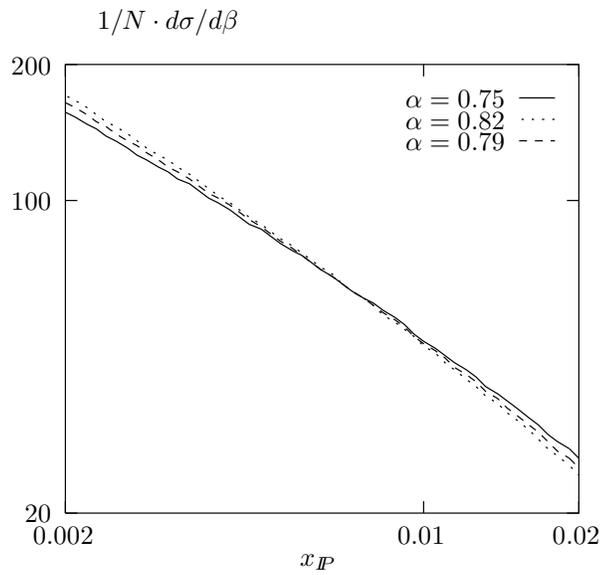
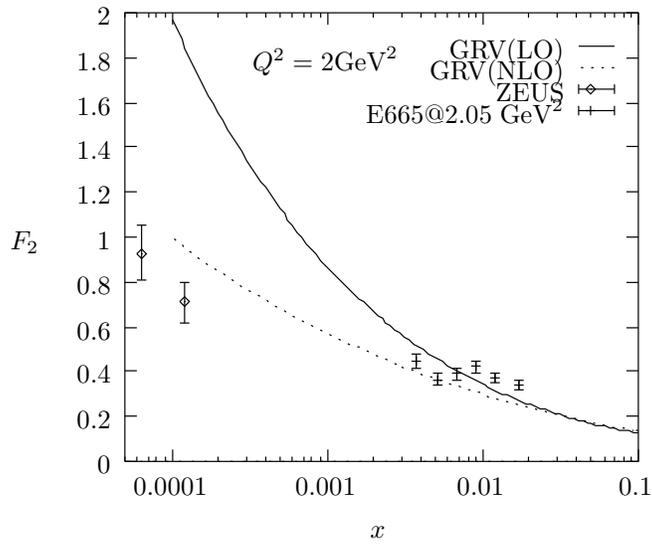
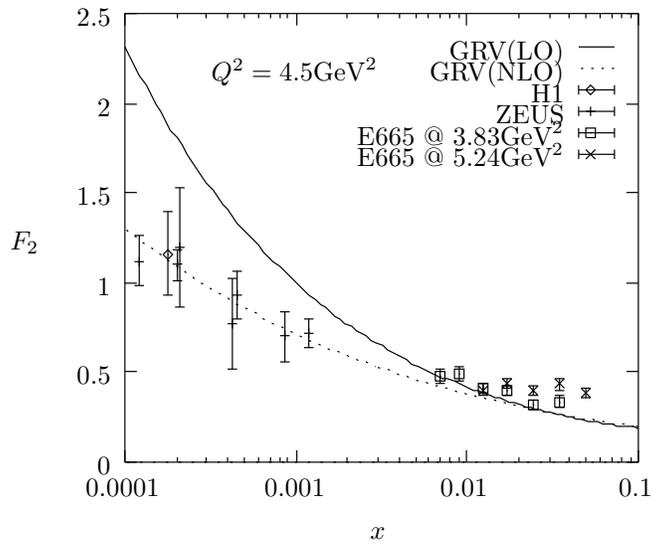
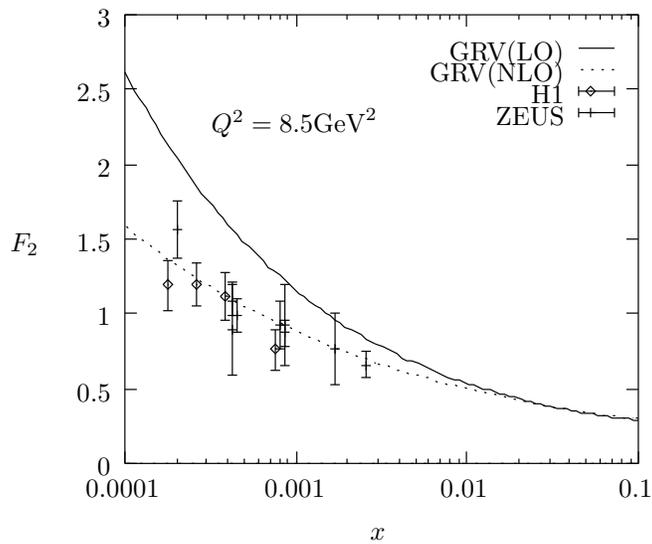
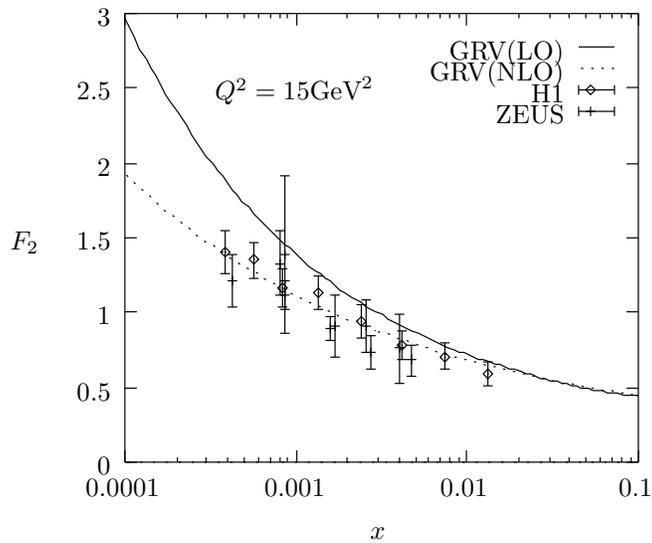
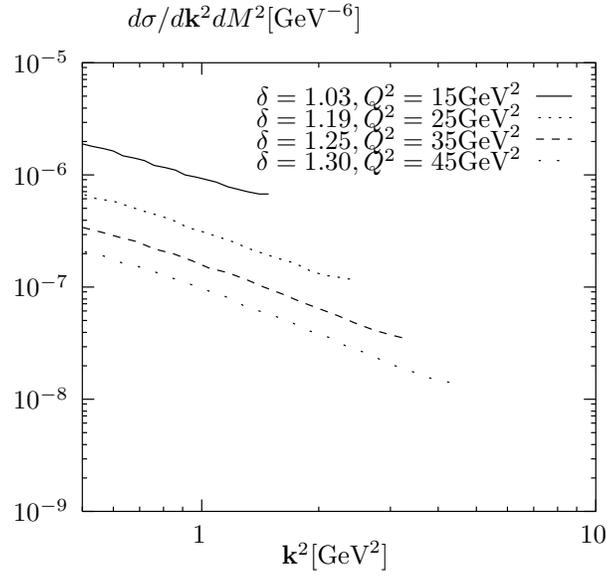
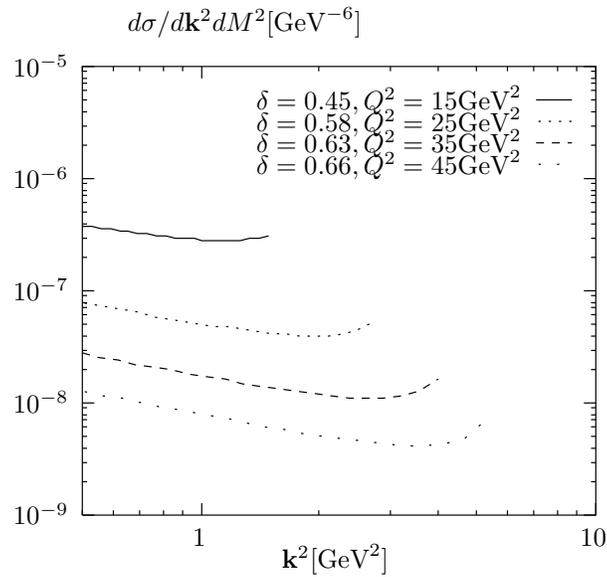
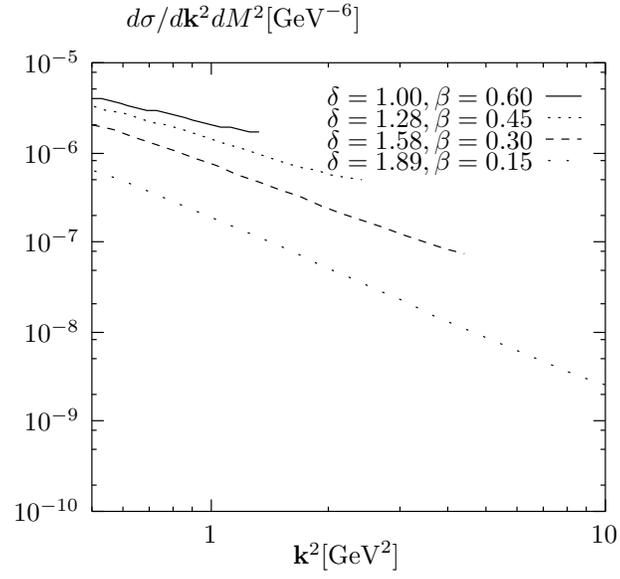
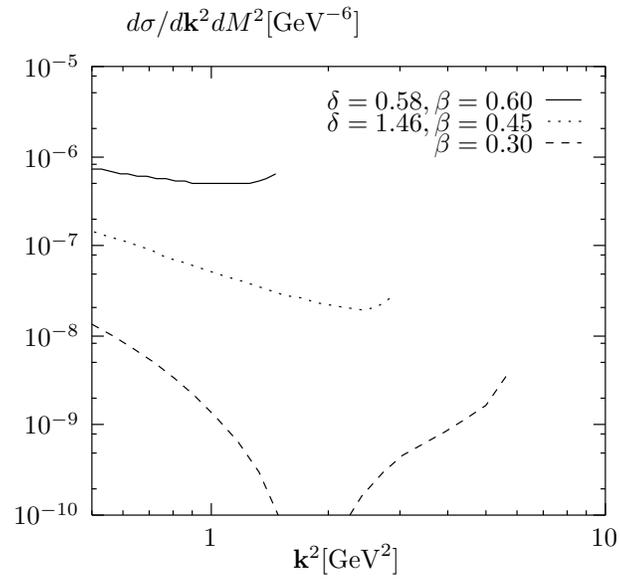
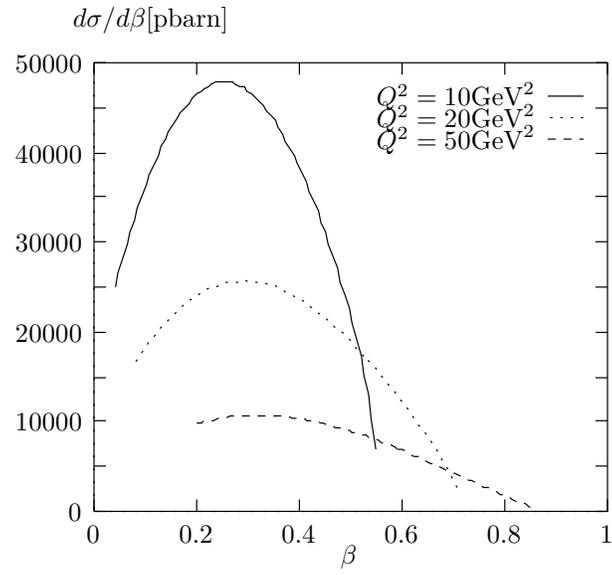
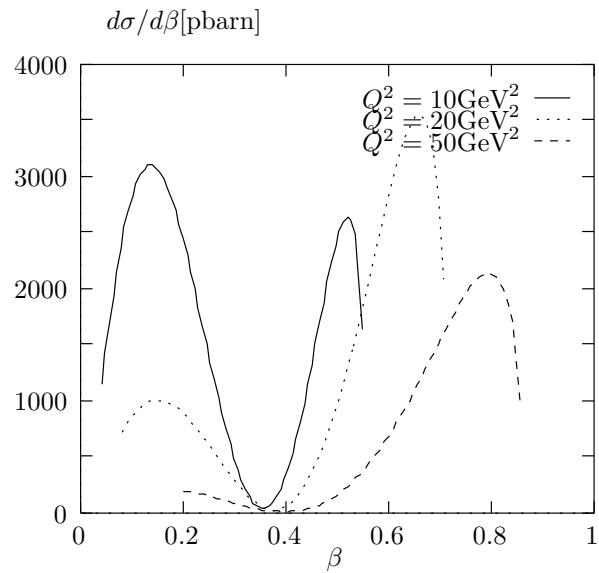
\end{document}

%% file: fig1.pstex_t
\begin{picture}(0,0)%
\epsfig{file=fig1.pstex}%
\end{picture}%
\setlength{\unitlength}{0.00075000in}%
\begingroup\makeatletter\ifx\SetFigFont\undefined
% extract first six characters in \fmtname
\def\x#1#2#3#4#5#6#7\relax{\def\x{#1#2#3#4#5#6}}%
\expandafter\x\fmtname xxxxxx\relax \def\y{splain}%
\ifx\x\y   % LaTeX or SliTeX?
\gdef\SetFigFont#1#2#3{%
  \ifnum #1<17\tiny\else \ifnum #1<20\small\else
  \ifnum #1<24\normalsize\else \ifnum #1<29\large\else
  \ifnum #1<34\Large\else \ifnum #1<41\LARGE\else
     \huge\fi\fi\fi\fi\fi\fi
  \csname #3\endcsname}%
\else
\gdef\SetFigFont#1#2#3{\begingroup
  \count@#1\relax \ifnum 25<\count@\count@25\fi
  \def\x{\endgroup\@setsize\SetFigFont{#2pt}}%
  \expandafter\x
    \csname \romannumeral\the\count@ pt\expandafter\endcsname
    \csname @\romannumeral\the\count@ pt\endcsname
  \csname #3\endcsname}%
\fi
\fi\endgroup
\begin{picture}(2850,2274)(3301,-2623)
\put(3301,-886){\makebox(0,0)[lb]{\smash{\SetFigFont{11}{13.2}{rm}$q$}}}
\put(6151,-586){\makebox(0,0)[lb]{\smash{\SetFigFont{11}{13.2}{rm}$k$}}}
\put(3601,-2311){\makebox(0,0)[lb]{\smash{\SetFigFont{11}{13.2}{rm}$p$}}}
\put(4426,-1486){\makebox(0,0)[lb]{\smash{\SetFigFont{11}{13.2}{rm}$l$}}}
\put(5401,-1486){\makebox(0,0)[lb]{\smash{\SetFigFont{11}{13.2}{rm}$-l+x_{\Pam}p$}}}
\put(6151,-2311){\makebox(0,0)[lb]{\smash{\SetFigFont{11}{13.2}{rm}$p-x_{\Pam}p$}}}
\put(6151,-1111){\makebox(0,0)[lb]{\smash{\SetFigFont{11}{13.2}{rm}$q-k+x_{\Pam}p$}}}
\end{picture}

%% file: fig2a.pstex_t
\begin{picture}(0,0)%
\epsfig{file=fig2a.pstex}%
\end{picture}%
\setlength{\unitlength}{0.01250000in}%
\begingroup\makeatletter\ifx\SetFigFont\undefined
% extract first six characters in \fmtname
\def\x#1#2#3#4#5#6#7\relax{\def\x{#1#2#3#4#5#6}}%
\expandafter\x\fmtname xxxxxx\relax \def\y{splain}%
\ifx\x\y   % LaTeX or SliTeX?
\gdef\SetFigFont#1#2#3{%
  \ifnum #1<17\tiny\else \ifnum #1<20\small\else
  \ifnum #1<24\normalsize\else \ifnum #1<29\large\else
  \ifnum #1<34\Large\else \ifnum #1<41\LARGE\else
     \huge\fi\fi\fi\fi\fi\fi
  \csname #3\endcsname}%
\else
\gdef\SetFigFont#1#2#3{\begingroup
  \count@#1\relax \ifnum 25<\count@\count@25\fi
  \def\x{\endgroup\@setsize\SetFigFont{#2pt}}%
  \expandafter\x
    \csname \romannumeral\the\count@ pt\expandafter\endcsname
    \csname @\romannumeral\the\count@ pt\endcsname
  \csname #3\endcsname}%
\fi
\fi\endgroup
\begin{picture}(268,234)(28,579)
\put( 95,805){\makebox(0,0)[b]{\smash{\SetFigFont{10}{12.0}{rm}$d \sigma/d \beta [\mbox{pbarn}]$}}}
\put( 66,598){\makebox(0,0)[rb]{\smash{\SetFigFont{10}{12.0}{rm}0}}}
\put( 66,629){\makebox(0,0)[rb]{\smash{\SetFigFont{10}{12.0}{rm}5000}}}
\put( 66,661){\makebox(0,0)[rb]{\smash{\SetFigFont{10}{12.0}{rm}10000}}}
\put( 66,692){\makebox(0,0)[rb]{\smash{\SetFigFont{10}{12.0}{rm}15000}}}
\put( 66,723){\makebox(0,0)[rb]{\smash{\SetFigFont{10}{12.0}{rm}20000}}}
\put( 66,755){\makebox(0,0)[rb]{\smash{\SetFigFont{10}{12.0}{rm}25000}}}
\put( 66,786){\makebox(0,0)[rb]{\smash{\SetFigFont{10}{12.0}{rm}30000}}}
\put( 71,589){\makebox(0,0)[b]{\smash{\SetFigFont{10}{12.0}{rm}0.001}}}
\put(286,589){\makebox(0,0)[b]{\smash{\SetFigFont{10}{12.0}{rm}0.01}}}
\put(178,580){\makebox(0,0)[b]{\smash{\SetFigFont{10}{12.0}{rm}$x_{\Pam}$}}}
\put(256,772){\makebox(0,0)[rb]{\smash{\SetFigFont{10}{12.0}{rm}Hybrid}}}
\put(256,763){\makebox(0,0)[rb]{\smash{\SetFigFont{10}{12.0}{rm}GRV(L)}}}
\put(256,754){\makebox(0,0)[rb]{\smash{\SetFigFont{10}{12.0}{rm}GRV(NL)}}}
\put(256,745){\makebox(0,0)[rb]{\smash{\SetFigFont{10}{12.0}{rm}soft Pomeron}}}
\end{picture}

%% file: fig2b.pstex_t
\begin{picture}(0,0)%
\epsfig{file=fig2b.pstex}%
\end{picture}%
\setlength{\unitlength}{0.01250000in}%
\begingroup\makeatletter\ifx\SetFigFont\undefined
% extract first six characters in \fmtname
\def\x#1#2#3#4#5#6#7\relax{\def\x{#1#2#3#4#5#6}}%
\expandafter\x\fmtname xxxxxx\relax \def\y{splain}%
\ifx\x\y   % LaTeX or SliTeX?
\gdef\SetFigFont#1#2#3{%
  \ifnum #1<17\tiny\else \ifnum #1<20\small\else
  \ifnum #1<24\normalsize\else \ifnum #1<29\large\else
  \ifnum #1<34\Large\else \ifnum #1<41\LARGE\else
     \huge\fi\fi\fi\fi\fi\fi
  \csname #3\endcsname}%
\else
\gdef\SetFigFont#1#2#3{\begingroup
  \count@#1\relax \ifnum 25<\count@\count@25\fi
  \def\x{\endgroup\@setsize\SetFigFont{#2pt}}%
  \expandafter\x
    \csname \romannumeral\the\count@ pt\expandafter\endcsname
    \csname @\romannumeral\the\count@ pt\endcsname
  \csname #3\endcsname}%
\fi
\fi\endgroup
\begin{picture}(268,234)(28,579)
\put( 95,805){\makebox(0,0)[b]{\smash{\SetFigFont{10}{12.0}{rm}$d \sigma/d \beta [\mbox{pbarn}]$}}}
\put( 66,598){\makebox(0,0)[rb]{\smash{\SetFigFont{10}{12.0}{rm}0}}}
\put( 66,629){\makebox(0,0)[rb]{\smash{\SetFigFont{10}{12.0}{rm}1000}}}
\put( 66,661){\makebox(0,0)[rb]{\smash{\SetFigFont{10}{12.0}{rm}2000}}}
\put( 66,692){\makebox(0,0)[rb]{\smash{\SetFigFont{10}{12.0}{rm}3000}}}
\put( 66,723){\makebox(0,0)[rb]{\smash{\SetFigFont{10}{12.0}{rm}4000}}}
\put( 66,755){\makebox(0,0)[rb]{\smash{\SetFigFont{10}{12.0}{rm}5000}}}
\put( 66,786){\makebox(0,0)[rb]{\smash{\SetFigFont{10}{12.0}{rm}6000}}}
\put( 71,589){\makebox(0,0)[b]{\smash{\SetFigFont{10}{12.0}{rm}0.001}}}
\put(286,589){\makebox(0,0)[b]{\smash{\SetFigFont{10}{12.0}{rm}0.01}}}
\put(178,580){\makebox(0,0)[b]{\smash{\SetFigFont{10}{12.0}{rm}$x_{\Pam}$}}}
\put(256,772){\makebox(0,0)[rb]{\smash{\SetFigFont{10}{12.0}{rm}Hybrid}}}
\put(256,763){\makebox(0,0)[rb]{\smash{\SetFigFont{10}{12.0}{rm}GRV(L)}}}
\put(256,754){\makebox(0,0)[rb]{\smash{\SetFigFont{10}{12.0}{rm}GRV(NL)}}}
\put(256,745){\makebox(0,0)[rb]{\smash{\SetFigFont{10}{12.0}{rm}soft Pomeron}}}
\end{picture}

%% file: fig2c.pstex_t
\begin{picture}(0,0)%
\epsfig{file=fig2c.pstex}%
\end{picture}%
\setlength{\unitlength}{0.01250000in}%
\begingroup\makeatletter\ifx\SetFigFont\undefined
% extract first six characters in \fmtname
\def\x#1#2#3#4#5#6#7\relax{\def\x{#1#2#3#4#5#6}}%
\expandafter\x\fmtname xxxxxx\relax \def\y{splain}%
\ifx\x\y   % LaTeX or SliTeX?
\gdef\SetFigFont#1#2#3{%
  \ifnum #1<17\tiny\else \ifnum #1<20\small\else
  \ifnum #1<24\normalsize\else \ifnum #1<29\large\else
  \ifnum #1<34\Large\else \ifnum #1<41\LARGE\else
     \huge\fi\fi\fi\fi\fi\fi
  \csname #3\endcsname}%
\else
\gdef\SetFigFont#1#2#3{\begingroup
  \count@#1\relax \ifnum 25<\count@\count@25\fi
  \def\x{\endgroup\@setsize\SetFigFont{#2pt}}%
  \expandafter\x
    \csname \romannumeral\the\count@ pt\expandafter\endcsname
    \csname @\romannumeral\the\count@ pt\endcsname
  \csname #3\endcsname}%
\fi
\fi\endgroup
\begin{picture}(274,234)(22,579)
\put(115,805){\makebox(0,0)[b]{\smash{\SetFigFont{10}{12.0}{rm}$1/N \cdot d \sigma/d \beta $}}}
\put( 66,598){\makebox(0,0)[rb]{\smash{\SetFigFont{10}{12.0}{rm}20}}}
\put( 66,729){\makebox(0,0)[rb]{\smash{\SetFigFont{10}{12.0}{rm}100}}}
\put( 66,786){\makebox(0,0)[rb]{\smash{\SetFigFont{10}{12.0}{rm}200}}}
\put( 71,589){\makebox(0,0)[b]{\smash{\SetFigFont{10}{12.0}{rm}0.002}}}
\put(221,589){\makebox(0,0)[b]{\smash{\SetFigFont{10}{12.0}{rm}0.01}}}
\put(286,589){\makebox(0,0)[b]{\smash{\SetFigFont{10}{12.0}{rm}0.02}}}
\put(178,580){\makebox(0,0)[b]{\smash{\SetFigFont{10}{12.0}{rm}$x_{\Pam}$}}}
\put(256,772){\makebox(0,0)[rb]{\smash{\SetFigFont{10}{12.0}{rm}$\alpha=0.75$}}}
\put(256,763){\makebox(0,0)[rb]{\smash{\SetFigFont{10}{12.0}{rm}$\alpha=0.82$}}}
\put(256,754){\makebox(0,0)[rb]{\smash{\SetFigFont{10}{12.0}{rm}$\alpha=0.79$}}}
\end{picture}

%% file: at2.pstex_t
\begin{picture}(0,0)%
\epsfig{file=at2.pstex}%
\end{picture}%
\setlength{\unitlength}{0.01250000in}%
\begingroup\makeatletter\ifx\SetFigFont\undefined
% extract first six characters in \fmtname
\def\x#1#2#3#4#5#6#7\relax{\def\x{#1#2#3#4#5#6}}%
\expandafter\x\fmtname xxxxxx\relax \def\y{splain}%
\ifx\x\y   % LaTeX or SliTeX?
\gdef\SetFigFont#1#2#3{%
  \ifnum #1<17\tiny\else \ifnum #1<20\small\else
  \ifnum #1<24\normalsize\else \ifnum #1<29\large\else
  \ifnum #1<34\Large\else \ifnum #1<41\LARGE\else
     \huge\fi\fi\fi\fi\fi\fi
  \csname #3\endcsname}%
\else
\gdef\SetFigFont#1#2#3{\begingroup
  \count@#1\relax \ifnum 25<\count@\count@25\fi
  \def\x{\endgroup\@setsize\SetFigFont{#2pt}}%
  \expandafter\x
    \csname \romannumeral\the\count@ pt\expandafter\endcsname
    \csname @\romannumeral\the\count@ pt\endcsname
  \csname #3\endcsname}%
\fi
\fi\endgroup
\begin{picture}(271,224)(22,570)
\put(256,745){\makebox(0,0)[rb]{\smash{\SetFigFont{10}{12.0}{rm}E665@2.05 $\mbox{GeV}^2$}}}
\put( 66,598){\makebox(0,0)[rb]{\smash{\SetFigFont{10}{12.0}{rm}0}}}
\put( 66,617){\makebox(0,0)[rb]{\smash{\SetFigFont{10}{12.0}{rm}0.2}}}
\put( 66,636){\makebox(0,0)[rb]{\smash{\SetFigFont{10}{12.0}{rm}0.4}}}
\put( 66,654){\makebox(0,0)[rb]{\smash{\SetFigFont{10}{12.0}{rm}0.6}}}
\put( 66,673){\makebox(0,0)[rb]{\smash{\SetFigFont{10}{12.0}{rm}0.8}}}
\put( 66,692){\makebox(0,0)[rb]{\smash{\SetFigFont{10}{12.0}{rm}1}}}
\put( 66,711){\makebox(0,0)[rb]{\smash{\SetFigFont{10}{12.0}{rm}1.2}}}
\put( 66,730){\makebox(0,0)[rb]{\smash{\SetFigFont{10}{12.0}{rm}1.4}}}
\put( 66,748){\makebox(0,0)[rb]{\smash{\SetFigFont{10}{12.0}{rm}1.6}}}
\put( 66,767){\makebox(0,0)[rb]{\smash{\SetFigFont{10}{12.0}{rm}1.8}}}
\put( 66,786){\makebox(0,0)[rb]{\smash{\SetFigFont{10}{12.0}{rm}2}}}
\put( 91,589){\makebox(0,0)[b]{\smash{\SetFigFont{10}{12.0}{rm}0.0001}}}
\put(156,589){\makebox(0,0)[b]{\smash{\SetFigFont{10}{12.0}{rm}0.001}}}
\put(221,589){\makebox(0,0)[b]{\smash{\SetFigFont{10}{12.0}{rm}0.01}}}
\put(286,589){\makebox(0,0)[b]{\smash{\SetFigFont{10}{12.0}{rm}0.1}}}
\put( 30,692){\makebox(0,0)[b]{\smash{\SetFigFont{10}{12.0}{rm}$F_2$}}}
\put(178,571){\makebox(0,0)[b]{\smash{\SetFigFont{10}{12.0}{rm}$x$}}}
\put(156,767){\makebox(0,0)[b]{\smash{\SetFigFont{10}{12.0}{rm}$Q^2 = 2 \mbox{GeV}^2$}}}
\put(256,772){\makebox(0,0)[rb]{\smash{\SetFigFont{10}{12.0}{rm}GRV(LO)}}}
\put(256,763){\makebox(0,0)[rb]{\smash{\SetFigFont{10}{12.0}{rm}GRV(NLO)}}}
\put(256,754){\makebox(0,0)[rb]{\smash{\SetFigFont{10}{12.0}{rm}ZEUS}}}
\end{picture}

%% file: at4.5.pstex_t
\begin{picture}(0,0)%
\epsfig{file=at4.5.pstex}%
\end{picture}%
\setlength{\unitlength}{0.01250000in}%
\begingroup\makeatletter\ifx\SetFigFont\undefined
% extract first six characters in \fmtname
\def\x#1#2#3#4#5#6#7\relax{\def\x{#1#2#3#4#5#6}}%
\expandafter\x\fmtname xxxxxx\relax \def\y{splain}%
\ifx\x\y   % LaTeX or SliTeX?
\gdef\SetFigFont#1#2#3{%
  \ifnum #1<17\tiny\else \ifnum #1<20\small\else
  \ifnum #1<24\normalsize\else \ifnum #1<29\large\else
  \ifnum #1<34\Large\else \ifnum #1<41\LARGE\else
     \huge\fi\fi\fi\fi\fi\fi
  \csname #3\endcsname}%
\else
\gdef\SetFigFont#1#2#3{\begingroup
  \count@#1\relax \ifnum 25<\count@\count@25\fi
  \def\x{\endgroup\@setsize\SetFigFont{#2pt}}%
  \expandafter\x
    \csname \romannumeral\the\count@ pt\expandafter\endcsname
    \csname @\romannumeral\the\count@ pt\endcsname
  \csname #3\endcsname}%
\fi
\fi\endgroup
\begin{picture}(271,224)(22,570)
\put(256,727){\makebox(0,0)[rb]{\smash{\SetFigFont{10}{12.0}{rm}E665 @ $5.24 \mbox{GeV}^2$}}}
\put( 66,598){\makebox(0,0)[rb]{\smash{\SetFigFont{10}{12.0}{rm}0}}}
\put( 66,636){\makebox(0,0)[rb]{\smash{\SetFigFont{10}{12.0}{rm}0.5}}}
\put( 66,673){\makebox(0,0)[rb]{\smash{\SetFigFont{10}{12.0}{rm}1}}}
\put( 66,711){\makebox(0,0)[rb]{\smash{\SetFigFont{10}{12.0}{rm}1.5}}}
\put( 66,748){\makebox(0,0)[rb]{\smash{\SetFigFont{10}{12.0}{rm}2}}}
\put( 66,786){\makebox(0,0)[rb]{\smash{\SetFigFont{10}{12.0}{rm}2.5}}}
\put( 71,589){\makebox(0,0)[b]{\smash{\SetFigFont{10}{12.0}{rm}0.0001}}}
\put(143,589){\makebox(0,0)[b]{\smash{\SetFigFont{10}{12.0}{rm}0.001}}}
\put(214,589){\makebox(0,0)[b]{\smash{\SetFigFont{10}{12.0}{rm}0.01}}}
\put(286,589){\makebox(0,0)[b]{\smash{\SetFigFont{10}{12.0}{rm}0.1}}}
\put( 30,692){\makebox(0,0)[b]{\smash{\SetFigFont{10}{12.0}{rm}$F_2$}}}
\put(178,571){\makebox(0,0)[b]{\smash{\SetFigFont{10}{12.0}{rm}$x$}}}
\put(143,763){\makebox(0,0)[b]{\smash{\SetFigFont{10}{12.0}{rm}$Q^2 = 4.5 \mbox{GeV}^2$}}}
\put(256,772){\makebox(0,0)[rb]{\smash{\SetFigFont{10}{12.0}{rm}GRV(LO)}}}
\put(256,763){\makebox(0,0)[rb]{\smash{\SetFigFont{10}{12.0}{rm}GRV(NLO)}}}
\put(256,754){\makebox(0,0)[rb]{\smash{\SetFigFont{10}{12.0}{rm}H1}}}
\put(256,745){\makebox(0,0)[rb]{\smash{\SetFigFont{10}{12.0}{rm}ZEUS}}}
\put(256,736){\makebox(0,0)[rb]{\smash{\SetFigFont{10}{12.0}{rm}E665 @ $3.83 \mbox{GeV}^2$}}}
\end{picture}

%% file: at8.5.pstex_t
\begin{picture}(0,0)%
\epsfig{file=at8.5.pstex}%
\end{picture}%
\setlength{\unitlength}{0.01250000in}%
\begingroup\makeatletter\ifx\SetFigFont\undefined
% extract first six characters in \fmtname
\def\x#1#2#3#4#5#6#7\relax{\def\x{#1#2#3#4#5#6}}%
\expandafter\x\fmtname xxxxxx\relax \def\y{splain}%
\ifx\x\y   % LaTeX or SliTeX?
\gdef\SetFigFont#1#2#3{%
  \ifnum #1<17\tiny\else \ifnum #1<20\small\else
  \ifnum #1<24\normalsize\else \ifnum #1<29\large\else
  \ifnum #1<34\Large\else \ifnum #1<41\LARGE\else
     \huge\fi\fi\fi\fi\fi\fi
  \csname #3\endcsname}%
\else
\gdef\SetFigFont#1#2#3{\begingroup
  \count@#1\relax \ifnum 25<\count@\count@25\fi
  \def\x{\endgroup\@setsize\SetFigFont{#2pt}}%
  \expandafter\x
    \csname \romannumeral\the\count@ pt\expandafter\endcsname
    \csname @\romannumeral\the\count@ pt\endcsname
  \csname #3\endcsname}%
\fi
\fi\endgroup
\begin{picture}(271,224)(22,570)
\put(256,745){\makebox(0,0)[rb]{\smash{\SetFigFont{10}{12.0}{rm}ZEUS}}}
\put( 66,598){\makebox(0,0)[rb]{\smash{\SetFigFont{10}{12.0}{rm}0}}}
\put( 66,629){\makebox(0,0)[rb]{\smash{\SetFigFont{10}{12.0}{rm}0.5}}}
\put( 66,661){\makebox(0,0)[rb]{\smash{\SetFigFont{10}{12.0}{rm}1}}}
\put( 66,692){\makebox(0,0)[rb]{\smash{\SetFigFont{10}{12.0}{rm}1.5}}}
\put( 66,723){\makebox(0,0)[rb]{\smash{\SetFigFont{10}{12.0}{rm}2}}}
\put( 66,755){\makebox(0,0)[rb]{\smash{\SetFigFont{10}{12.0}{rm}2.5}}}
\put( 66,786){\makebox(0,0)[rb]{\smash{\SetFigFont{10}{12.0}{rm}3}}}
\put( 71,589){\makebox(0,0)[b]{\smash{\SetFigFont{10}{12.0}{rm}0.0001}}}
\put(143,589){\makebox(0,0)[b]{\smash{\SetFigFont{10}{12.0}{rm}0.001}}}
\put(214,589){\makebox(0,0)[b]{\smash{\SetFigFont{10}{12.0}{rm}0.01}}}
\put(286,589){\makebox(0,0)[b]{\smash{\SetFigFont{10}{12.0}{rm}0.1}}}
\put( 30,692){\makebox(0,0)[b]{\smash{\SetFigFont{10}{12.0}{rm}$F_2$}}}
\put(178,571){\makebox(0,0)[b]{\smash{\SetFigFont{10}{12.0}{rm}$x$}}}
\put(143,742){\makebox(0,0)[b]{\smash{\SetFigFont{10}{12.0}{rm}$Q^2 = 8.5 \mbox{GeV}^2$}}}
\put(256,772){\makebox(0,0)[rb]{\smash{\SetFigFont{10}{12.0}{rm}GRV(LO)}}}
\put(256,763){\makebox(0,0)[rb]{\smash{\SetFigFont{10}{12.0}{rm}GRV(NLO)}}}
\put(256,754){\makebox(0,0)[rb]{\smash{\SetFigFont{10}{12.0}{rm}H1}}}
\end{picture}

%% file: at15.pstex_t
\begin{picture}(0,0)%
\epsfig{file=at15.pstex}%
\end{picture}%
\setlength{\unitlength}{0.01250000in}%
\begingroup\makeatletter\ifx\SetFigFont\undefined
% extract first six characters in \fmtname
\def\x#1#2#3#4#5#6#7\relax{\def\x{#1#2#3#4#5#6}}%
\expandafter\x\fmtname xxxxxx\relax \def\y{splain}%
\ifx\x\y   % LaTeX or SliTeX?
\gdef\SetFigFont#1#2#3{%
  \ifnum #1<17\tiny\else \ifnum #1<20\small\else
  \ifnum #1<24\normalsize\else \ifnum #1<29\large\else
  \ifnum #1<34\Large\else \ifnum #1<41\LARGE\else
     \huge\fi\fi\fi\fi\fi\fi
  \csname #3\endcsname}%
\else
\gdef\SetFigFont#1#2#3{\begingroup
  \count@#1\relax \ifnum 25<\count@\count@25\fi
  \def\x{\endgroup\@setsize\SetFigFont{#2pt}}%
  \expandafter\x
    \csname \romannumeral\the\count@ pt\expandafter\endcsname
    \csname @\romannumeral\the\count@ pt\endcsname
  \csname #3\endcsname}%
\fi
\fi\endgroup
\begin{picture}(271,224)(22,570)
\put(256,745){\makebox(0,0)[rb]{\smash{\SetFigFont{10}{12.0}{rm}ZEUS}}}
\put( 66,598){\makebox(0,0)[rb]{\smash{\SetFigFont{10}{12.0}{rm}0}}}
\put( 66,629){\makebox(0,0)[rb]{\smash{\SetFigFont{10}{12.0}{rm}0.5}}}
\put( 66,661){\makebox(0,0)[rb]{\smash{\SetFigFont{10}{12.0}{rm}1}}}
\put( 66,692){\makebox(0,0)[rb]{\smash{\SetFigFont{10}{12.0}{rm}1.5}}}
\put( 66,723){\makebox(0,0)[rb]{\smash{\SetFigFont{10}{12.0}{rm}2}}}
\put( 66,755){\makebox(0,0)[rb]{\smash{\SetFigFont{10}{12.0}{rm}2.5}}}
\put( 66,786){\makebox(0,0)[rb]{\smash{\SetFigFont{10}{12.0}{rm}3}}}
\put( 71,589){\makebox(0,0)[b]{\smash{\SetFigFont{10}{12.0}{rm}0.0001}}}
\put(143,589){\makebox(0,0)[b]{\smash{\SetFigFont{10}{12.0}{rm}0.001}}}
\put(214,589){\makebox(0,0)[b]{\smash{\SetFigFont{10}{12.0}{rm}0.01}}}
\put(286,589){\makebox(0,0)[b]{\smash{\SetFigFont{10}{12.0}{rm}0.1}}}
\put( 30,692){\makebox(0,0)[b]{\smash{\SetFigFont{10}{12.0}{rm}$F_2$}}}
\put(178,571){\makebox(0,0)[b]{\smash{\SetFigFont{10}{12.0}{rm}$x$}}}
\put(143,755){\makebox(0,0)[b]{\smash{\SetFigFont{10}{12.0}{rm}$Q^2 = 15 \mbox{GeV}^2$}}}
\put(256,772){\makebox(0,0)[rb]{\smash{\SetFigFont{10}{12.0}{rm}GRV(LO)}}}
\put(256,763){\makebox(0,0)[rb]{\smash{\SetFigFont{10}{12.0}{rm}GRV(NLO)}}}
\put(256,754){\makebox(0,0)[rb]{\smash{\SetFigFont{10}{12.0}{rm}H1}}}
\end{picture}

%% file: fig4a.pstex_t
\begin{picture}(0,0)%
\epsfig{file=fig4a.pstex}%
\end{picture}%
\setlength{\unitlength}{0.01250000in}%
\begingroup\makeatletter\ifx\SetFigFont\undefined
% extract first six characters in \fmtname
\def\x#1#2#3#4#5#6#7\relax{\def\x{#1#2#3#4#5#6}}%
\expandafter\x\fmtname xxxxxx\relax \def\y{splain}%
\ifx\x\y   % LaTeX or SliTeX?
\gdef\SetFigFont#1#2#3{%
  \ifnum #1<17\tiny\else \ifnum #1<20\small\else
  \ifnum #1<24\normalsize\else \ifnum #1<29\large\else
  \ifnum #1<34\Large\else \ifnum #1<41\LARGE\else
     \huge\fi\fi\fi\fi\fi\fi
  \csname #3\endcsname}%
\else
\gdef\SetFigFont#1#2#3{\begingroup
  \count@#1\relax \ifnum 25<\count@\count@25\fi
  \def\x{\endgroup\@setsize\SetFigFont{#2pt}}%
  \expandafter\x
    \csname \romannumeral\the\count@ pt\expandafter\endcsname
    \csname @\romannumeral\the\count@ pt\endcsname
  \csname #3\endcsname}%
\fi
\fi\endgroup
\begin{picture}(259,234)(32,579)
\put(140,805){\makebox(0,0)[b]{\smash{\SetFigFont{10}{12.0}{rm}$d \sigma/ d \kf^2 d M^2 [\mbox{GeV}^{-6}]$}}}
\put( 66,598){\makebox(0,0)[rb]{\smash{\SetFigFont{10}{12.0}{rm}$10^{-9}$}}}
\put( 66,645){\makebox(0,0)[rb]{\smash{\SetFigFont{10}{12.0}{rm}$10^{-8}$}}}
\put( 66,692){\makebox(0,0)[rb]{\smash{\SetFigFont{10}{12.0}{rm}$10^{-7}$}}}
\put( 66,739){\makebox(0,0)[rb]{\smash{\SetFigFont{10}{12.0}{rm}$10^{-6}$}}}
\put( 66,786){\makebox(0,0)[rb]{\smash{\SetFigFont{10}{12.0}{rm}$10^{-5}$}}}
\put(121,589){\makebox(0,0)[b]{\smash{\SetFigFont{10}{12.0}{rm}1}}}
\put(286,589){\makebox(0,0)[b]{\smash{\SetFigFont{10}{12.0}{rm}10}}}
\put(178,580){\makebox(0,0)[b]{\smash{\SetFigFont{10}{12.0}{rm}$\kf^2 [\mbox{GeV}^2]$}}}
\put(256,772){\makebox(0,0)[rb]{\smash{\SetFigFont{10}{12.0}{rm}$\delta=1.03,Q^2=15 \mbox{GeV}^2$}}}
\put(256,763){\makebox(0,0)[rb]{\smash{\SetFigFont{10}{12.0}{rm}$\delta=1.19,Q^2=25 \mbox{GeV}^2$}}}
\put(256,754){\makebox(0,0)[rb]{\smash{\SetFigFont{10}{12.0}{rm}$\delta=1.25,Q^2=35 \mbox{GeV}^2$}}}
\put(256,745){\makebox(0,0)[rb]{\smash{\SetFigFont{10}{12.0}{rm}$\delta=1.30,Q^2=45 \mbox{GeV}^2$}}}
\end{picture}

%% file: fig4b.pstex_t
\begin{picture}(0,0)%
\epsfig{file=fig4b.pstex}%
\end{picture}%
\setlength{\unitlength}{0.01250000in}%
\begingroup\makeatletter\ifx\SetFigFont\undefined
% extract first six characters in \fmtname
\def\x#1#2#3#4#5#6#7\relax{\def\x{#1#2#3#4#5#6}}%
\expandafter\x\fmtname xxxxxx\relax \def\y{splain}%
\ifx\x\y   % LaTeX or SliTeX?
\gdef\SetFigFont#1#2#3{%
  \ifnum #1<17\tiny\else \ifnum #1<20\small\else
  \ifnum #1<24\normalsize\else \ifnum #1<29\large\else
  \ifnum #1<34\Large\else \ifnum #1<41\LARGE\else
     \huge\fi\fi\fi\fi\fi\fi
  \csname #3\endcsname}%
\else
\gdef\SetFigFont#1#2#3{\begingroup
  \count@#1\relax \ifnum 25<\count@\count@25\fi
  \def\x{\endgroup\@setsize\SetFigFont{#2pt}}%
  \expandafter\x
    \csname \romannumeral\the\count@ pt\expandafter\endcsname
    \csname @\romannumeral\the\count@ pt\endcsname
  \csname #3\endcsname}%
\fi
\fi\endgroup
\begin{picture}(259,234)(32,579)
\put(140,805){\makebox(0,0)[b]{\smash{\SetFigFont{10}{12.0}{rm}$d \sigma/ d \kf^2 d M^2 [\mbox{GeV}^{-6}]$}}}
\put( 66,598){\makebox(0,0)[rb]{\smash{\SetFigFont{10}{12.0}{rm}$10^{-9}$}}}
\put( 66,645){\makebox(0,0)[rb]{\smash{\SetFigFont{10}{12.0}{rm}$10^{-8}$}}}
\put( 66,692){\makebox(0,0)[rb]{\smash{\SetFigFont{10}{12.0}{rm}$10^{-7}$}}}
\put( 66,739){\makebox(0,0)[rb]{\smash{\SetFigFont{10}{12.0}{rm}$10^{-6}$}}}
\put( 66,786){\makebox(0,0)[rb]{\smash{\SetFigFont{10}{12.0}{rm}$10^{-5}$}}}
\put(121,589){\makebox(0,0)[b]{\smash{\SetFigFont{10}{12.0}{rm}1}}}
\put(286,589){\makebox(0,0)[b]{\smash{\SetFigFont{10}{12.0}{rm}10}}}
\put(178,580){\makebox(0,0)[b]{\smash{\SetFigFont{10}{12.0}{rm}$\kf^2 [\mbox{GeV}^2]$}}}
\put(256,772){\makebox(0,0)[rb]{\smash{\SetFigFont{10}{12.0}{rm}$\delta=0.45,Q^2=15 \mbox{GeV}^2$}}}
\put(256,763){\makebox(0,0)[rb]{\smash{\SetFigFont{10}{12.0}{rm}$\delta=0.58,Q^2=25 \mbox{GeV}^2$}}}
\put(256,754){\makebox(0,0)[rb]{\smash{\SetFigFont{10}{12.0}{rm}$\delta=0.63,Q^2=35 \mbox{GeV}^2$}}}
\put(256,745){\makebox(0,0)[rb]{\smash{\SetFigFont{10}{12.0}{rm}$\delta=0.66,Q^2=45 \mbox{GeV}^2$}}}
\end{picture}

%% file: fig4c.pstex_t
\begin{picture}(0,0)%
\epsfig{file=fig4c.pstex}%
\end{picture}%
\setlength{\unitlength}{0.01250000in}%
\begingroup\makeatletter\ifx\SetFigFont\undefined
% extract first six characters in \fmtname
\def\x#1#2#3#4#5#6#7\relax{\def\x{#1#2#3#4#5#6}}%
\expandafter\x\fmtname xxxxxx\relax \def\y{splain}%
\ifx\x\y   % LaTeX or SliTeX?
\gdef\SetFigFont#1#2#3{%
  \ifnum #1<17\tiny\else \ifnum #1<20\small\else
  \ifnum #1<24\normalsize\else \ifnum #1<29\large\else
  \ifnum #1<34\Large\else \ifnum #1<41\LARGE\else
     \huge\fi\fi\fi\fi\fi\fi
  \csname #3\endcsname}%
\else
\gdef\SetFigFont#1#2#3{\begingroup
  \count@#1\relax \ifnum 25<\count@\count@25\fi
  \def\x{\endgroup\@setsize\SetFigFont{#2pt}}%
  \expandafter\x
    \csname \romannumeral\the\count@ pt\expandafter\endcsname
    \csname @\romannumeral\the\count@ pt\endcsname
  \csname #3\endcsname}%
\fi
\fi\endgroup
\begin{picture}(264,234)(27,579)
\put(135,805){\makebox(0,0)[b]{\smash{\SetFigFont{10}{12.0}{rm}$d \sigma/ d \kf^2 d M^2 [\mbox{GeV}^{-6}]$}}}
\put( 66,598){\makebox(0,0)[rb]{\smash{\SetFigFont{10}{12.0}{rm}$10^{-10}$}}}
\put( 66,636){\makebox(0,0)[rb]{\smash{\SetFigFont{10}{12.0}{rm}$10^{-9}$}}}
\put( 66,673){\makebox(0,0)[rb]{\smash{\SetFigFont{10}{12.0}{rm}$10^{-8}$}}}
\put( 66,711){\makebox(0,0)[rb]{\smash{\SetFigFont{10}{12.0}{rm}$10^{-7}$}}}
\put( 66,748){\makebox(0,0)[rb]{\smash{\SetFigFont{10}{12.0}{rm}$10^{-6}$}}}
\put( 66,786){\makebox(0,0)[rb]{\smash{\SetFigFont{10}{12.0}{rm}$10^{-5}$}}}
\put(121,589){\makebox(0,0)[b]{\smash{\SetFigFont{10}{12.0}{rm}1}}}
\put(286,589){\makebox(0,0)[b]{\smash{\SetFigFont{10}{12.0}{rm}10}}}
\put(178,580){\makebox(0,0)[b]{\smash{\SetFigFont{10}{12.0}{rm}$\kf^2 [\mbox{GeV}^2]$}}}
\put(256,772){\makebox(0,0)[rb]{\smash{\SetFigFont{10}{12.0}{rm}$\delta=1.00,\beta=0.60$}}}
\put(256,763){\makebox(0,0)[rb]{\smash{\SetFigFont{10}{12.0}{rm}$\delta=1.28,\beta=0.45$}}}
\put(256,754){\makebox(0,0)[rb]{\smash{\SetFigFont{10}{12.0}{rm}$\delta=1.58,\beta=0.30$}}}
\put(256,745){\makebox(0,0)[rb]{\smash{\SetFigFont{10}{12.0}{rm}$\delta=1.89,\beta=0.15$}}}
\end{picture}

%% file: fig4d.pstex_t
\begin{picture}(0,0)%
\epsfig{file=fig4d.pstex}%
\end{picture}%
\setlength{\unitlength}{0.01250000in}%
\begingroup\makeatletter\ifx\SetFigFont\undefined
% extract first six characters in \fmtname
\def\x#1#2#3#4#5#6#7\relax{\def\x{#1#2#3#4#5#6}}%
\expandafter\x\fmtname xxxxxx\relax \def\y{splain}%
\ifx\x\y   % LaTeX or SliTeX?
\gdef\SetFigFont#1#2#3{%
  \ifnum #1<17\tiny\else \ifnum #1<20\small\else
  \ifnum #1<24\normalsize\else \ifnum #1<29\large\else
  \ifnum #1<34\Large\else \ifnum #1<41\LARGE\else
     \huge\fi\fi\fi\fi\fi\fi
  \csname #3\endcsname}%
\else
\gdef\SetFigFont#1#2#3{\begingroup
  \count@#1\relax \ifnum 25<\count@\count@25\fi
  \def\x{\endgroup\@setsize\SetFigFont{#2pt}}%
  \expandafter\x
    \csname \romannumeral\the\count@ pt\expandafter\endcsname
    \csname @\romannumeral\the\count@ pt\endcsname
  \csname #3\endcsname}%
\fi
\fi\endgroup
\begin{picture}(264,234)(27,579)
\put(135,805){\makebox(0,0)[b]{\smash{\SetFigFont{10}{12.0}{rm}$d \sigma/ d \kf^2 d M^2 [\mbox{GeV}^{-6}]$}}}
\put( 66,598){\makebox(0,0)[rb]{\smash{\SetFigFont{10}{12.0}{rm}$10^{-10}$}}}
\put( 66,636){\makebox(0,0)[rb]{\smash{\SetFigFont{10}{12.0}{rm}$10^{-9}$}}}
\put( 66,673){\makebox(0,0)[rb]{\smash{\SetFigFont{10}{12.0}{rm}$10^{-8}$}}}
\put( 66,711){\makebox(0,0)[rb]{\smash{\SetFigFont{10}{12.0}{rm}$10^{-7}$}}}
\put( 66,748){\makebox(0,0)[rb]{\smash{\SetFigFont{10}{12.0}{rm}$10^{-6}$}}}
\put( 66,786){\makebox(0,0)[rb]{\smash{\SetFigFont{10}{12.0}{rm}$10^{-5}$}}}
\put(121,589){\makebox(0,0)[b]{\smash{\SetFigFont{10}{12.0}{rm}1}}}
\put(286,589){\makebox(0,0)[b]{\smash{\SetFigFont{10}{12.0}{rm}10}}}
\put(178,580){\makebox(0,0)[b]{\smash{\SetFigFont{10}{12.0}{rm}$\kf^2 [\mbox{GeV}^2]$}}}
\put(256,772){\makebox(0,0)[rb]{\smash{\SetFigFont{10}{12.0}{rm}$\delta=0.58,\beta=0.60$}}}
\put(256,763){\makebox(0,0)[rb]{\smash{\SetFigFont{10}{12.0}{rm}$\delta=1.46,\beta=0.45$}}}
\put(256,754){\makebox(0,0)[rb]{\smash{\SetFigFont{10}{12.0}{rm}$\beta=0.30$}}}
\end{picture}

%% file: fig5a.pstex_t
\begin{picture}(0,0)%
\epsfig{file=fig5a.pstex}%
\end{picture}%
\setlength{\unitlength}{0.01250000in}%
\begingroup\makeatletter\ifx\SetFigFont\undefined
% extract first six characters in \fmtname
\def\x#1#2#3#4#5#6#7\relax{\def\x{#1#2#3#4#5#6}}%
\expandafter\x\fmtname xxxxxx\relax \def\y{splain}%
\ifx\x\y   % LaTeX or SliTeX?
\gdef\SetFigFont#1#2#3{%
  \ifnum #1<17\tiny\else \ifnum #1<20\small\else
  \ifnum #1<24\normalsize\else \ifnum #1<29\large\else
  \ifnum #1<34\Large\else \ifnum #1<41\LARGE\else
     \huge\fi\fi\fi\fi\fi\fi
  \csname #3\endcsname}%
\else
\gdef\SetFigFont#1#2#3{\begingroup
  \count@#1\relax \ifnum 25<\count@\count@25\fi
  \def\x{\endgroup\@setsize\SetFigFont{#2pt}}%
  \expandafter\x
    \csname \romannumeral\the\count@ pt\expandafter\endcsname
    \csname @\romannumeral\the\count@ pt\endcsname
  \csname #3\endcsname}%
\fi
\fi\endgroup
\begin{picture}(260,234)(28,579)
\put( 95,805){\makebox(0,0)[b]{\smash{\SetFigFont{10}{12.0}{rm}$d \sigma/d \beta [\mbox{pbarn}]$}}}
\put( 66,598){\makebox(0,0)[rb]{\smash{\SetFigFont{10}{12.0}{rm}0}}}
\put( 66,636){\makebox(0,0)[rb]{\smash{\SetFigFont{10}{12.0}{rm}10000}}}
\put( 66,673){\makebox(0,0)[rb]{\smash{\SetFigFont{10}{12.0}{rm}20000}}}
\put( 66,711){\makebox(0,0)[rb]{\smash{\SetFigFont{10}{12.0}{rm}30000}}}
\put( 66,748){\makebox(0,0)[rb]{\smash{\SetFigFont{10}{12.0}{rm}40000}}}
\put( 66,786){\makebox(0,0)[rb]{\smash{\SetFigFont{10}{12.0}{rm}50000}}}
\put( 71,589){\makebox(0,0)[b]{\smash{\SetFigFont{10}{12.0}{rm}0}}}
\put(114,589){\makebox(0,0)[b]{\smash{\SetFigFont{10}{12.0}{rm}0.2}}}
\put(157,589){\makebox(0,0)[b]{\smash{\SetFigFont{10}{12.0}{rm}0.4}}}
\put(200,589){\makebox(0,0)[b]{\smash{\SetFigFont{10}{12.0}{rm}0.6}}}
\put(243,589){\makebox(0,0)[b]{\smash{\SetFigFont{10}{12.0}{rm}0.8}}}
\put(286,589){\makebox(0,0)[b]{\smash{\SetFigFont{10}{12.0}{rm}1}}}
\put(178,580){\makebox(0,0)[b]{\smash{\SetFigFont{10}{12.0}{rm}$\beta$}}}
\put(256,772){\makebox(0,0)[rb]{\smash{\SetFigFont{10}{12.0}{rm}$Q^2=10 \mbox{GeV}^2$}}}
\put(256,763){\makebox(0,0)[rb]{\smash{\SetFigFont{10}{12.0}{rm}$Q^2=20 \mbox{GeV}^2$}}}
\put(256,754){\makebox(0,0)[rb]{\smash{\SetFigFont{10}{12.0}{rm}$Q^2=50 \mbox{GeV}^2$}}}
\end{picture}

%% file: fig5b.pstex_t
\begin{picture}(0,0)%
\epsfig{file=fig5b.pstex}%
\end{picture}%
\setlength{\unitlength}{0.01250000in}%
\begingroup\makeatletter\ifx\SetFigFont\undefined
% extract first six characters in \fmtname
\def\x#1#2#3#4#5#6#7\relax{\def\x{#1#2#3#4#5#6}}%
\expandafter\x\fmtname xxxxxx\relax \def\y{splain}%
\ifx\x\y   % LaTeX or SliTeX?
\gdef\SetFigFont#1#2#3{%
  \ifnum #1<17\tiny\else \ifnum #1<20\small\else
  \ifnum #1<24\normalsize\else \ifnum #1<29\large\else
  \ifnum #1<34\Large\else \ifnum #1<41\LARGE\else
     \huge\fi\fi\fi\fi\fi\fi
  \csname #3\endcsname}%
\else
\gdef\SetFigFont#1#2#3{\begingroup
  \count@#1\relax \ifnum 25<\count@\count@25\fi
  \def\x{\endgroup\@setsize\SetFigFont{#2pt}}%
  \expandafter\x
    \csname \romannumeral\the\count@ pt\expandafter\endcsname
    \csname @\romannumeral\the\count@ pt\endcsname
  \csname #3\endcsname}%
\fi
\fi\endgroup
\begin{picture}(243,234)(45,579)
\put(115,805){\makebox(0,0)[b]{\smash{\SetFigFont{10}{12.0}{rm}$d \sigma/d \beta [\mbox{pbarn}]$}}}
\put( 66,598){\makebox(0,0)[rb]{\smash{\SetFigFont{10}{12.0}{rm}0}}}
\put( 66,645){\makebox(0,0)[rb]{\smash{\SetFigFont{10}{12.0}{rm}1000}}}
\put( 66,692){\makebox(0,0)[rb]{\smash{\SetFigFont{10}{12.0}{rm}2000}}}
\put( 66,739){\makebox(0,0)[rb]{\smash{\SetFigFont{10}{12.0}{rm}3000}}}
\put( 66,786){\makebox(0,0)[rb]{\smash{\SetFigFont{10}{12.0}{rm}4000}}}
\put( 71,589){\makebox(0,0)[b]{\smash{\SetFigFont{10}{12.0}{rm}0}}}
\put(114,589){\makebox(0,0)[b]{\smash{\SetFigFont{10}{12.0}{rm}0.2}}}
\put(157,589){\makebox(0,0)[b]{\smash{\SetFigFont{10}{12.0}{rm}0.4}}}
\put(200,589){\makebox(0,0)[b]{\smash{\SetFigFont{10}{12.0}{rm}0.6}}}
\put(243,589){\makebox(0,0)[b]{\smash{\SetFigFont{10}{12.0}{rm}0.8}}}
\put(286,589){\makebox(0,0)[b]{\smash{\SetFigFont{10}{12.0}{rm}1}}}
\put(178,580){\makebox(0,0)[b]{\smash{\SetFigFont{10}{12.0}{rm}$\beta$}}}
\put(256,772){\makebox(0,0)[rb]{\smash{\SetFigFont{10}{12.0}{rm}$Q^2=10 \mbox{GeV}^2$}}}
\put(256,763){\makebox(0,0)[rb]{\smash{\SetFigFont{10}{12.0}{rm}$Q^2=20 \mbox{GeV}^2$}}}
\put(256,754){\makebox(0,0)[rb]{\smash{\SetFigFont{10}{12.0}{rm}$Q^2=50 \mbox{GeV}^2$}}}
\end{picture}